# Identification of Ischemic Heart Disease by using machine learning technique based on parameters measuring Heart Rate Variability

Giulia Silveri
*Dep. Engineering and Architecture*
University of Trieste
Trieste, Italy
giulia.silveri@phd.units.it

Marco Merlo
*Cardiovascular Department*
University of Trieste
Trieste, Italy
marco.merlo@asuits.sanita.fvg.it

Luca Restivo
*Cardiovascular Department*
University of Trieste
Trieste, Italy
dott.lucarestivo92@gmail.com

Beatrice De Paola
*Dep. Engineering and Architecture*
University of Trieste
Trieste, Italy
depaolabeatrice@outlook.it

Aleksandar Miladinović
*Dep. Engineering and Architecture*
University of Trieste
Trieste, Italy
aleksandar.miladinovic@phd.units.it

Miloš Ajčević
*Dep. Engineering and Architecture*
University of Trieste
Trieste, Italy
majcevic@units.it

Gianfranco Sinagra
*Cardiovascular Department*
University of Trieste
Trieste, Italy
gianfranco.sinagra@asuits.sanita.fvg.it

Agostino Accardo
*Dep. Engineering and Architecture*
University of Trieste
Trieste, Italy
accardo@units.it

*Abstract*— The diagnosis of heart diseases is a difficult task generally addressed by an appropriate examination of patients' clinical data. Recently, the use of heart rate variability (HRV) analysis as well as of some machine learning algorithms, has proved to be a valuable support in the diagnosis process. However, till now, ischemic heart disease (IHD) has been diagnosed on the basis of Artificial Neural Networks (ANN) applied only to signs, symptoms and sequential ECG and coronary angiography, an invasive tool, while could be probably identified in a non-invasive way by using parameters extracted from HRV, a signal easily obtained from the ECG. In this study, 18 non-invasive features (age, gender, left ventricular ejection fraction and 15 obtained from HRV) of 243 subjects (156 normal subjects and 87 IHD patients) were used to train and validate a series of several ANN, different for number of input and hidden nodes. The best result was obtained using 7 input parameters and 7 hidden nodes with an accuracy of 98.9% and 82% for the training and validation dataset, respectively.

## I. INTRODUCTION

Ischemic heart disease (IHD) is a condition in which there is an inadequate supply of blood and oxygen to a portion of the myocardium. This condition typically occurs when there is an imbalance between myocardial oxygen supply and demand, typically due to atherosclerotic heart disease [1]. Patients are generally asymptomatic or exhibit no typical signs and symptoms until ischemic heart disease manifests itself as angina, myocardial infarction or sudden cardiac death [2].

Coronary angiography provides a definite diagnosis of IHD. However, coronary angiography is an invasive tool requiring the use of possibly toxic contrast means. Therefore, it is used in advanced stages of the disease, in presence of symptoms, or left ventricular dysfunction while, in preliminary stages, the electrocardiogram (ECG) could help in the diagnosis of suspect IHD. From ECG the variation over time of the period between consecutive heartbeats (HRV) that presents the autonomic nervous system and cardiovascular autonomic regulation could be extracted [3]. Patients with IHD generally show significant different values of linear [4] and non-linear [5] HRV parameters than normal subjects.

Over the past years, artificial intelligence and machine learning techniques constituted a predictive and powerful instrument for identifying cardiovascular diseases. Acharya et al. [6] have developed a disease classification system by using an artificial neural network (ANN) in order to identify eight cardiac disorders (not considering ischemic and dilated cardiomyopathy separately). The MIT-BIH arrhythmia database data were used and three non-linear features were selected: spectral entropy, Poincaré plot geometry and largest Lyapunov exponent, obtaining an accuracy level of 80-85%. In addition, Obbaya et al. [7] developed a multilayer feed forward neural network to discriminate normal subjects from patients suffering from either congestive heart failure or myocardial infarction disease. They used HRV data of PhysioBank Interbeat (RR) Interval Database, considering separately features extracted from HRV in the time domain (five parameters), in frequency domain (three parameters) and six non-linear parameters, obtaining an average classification rate of 93%, 96% and 97% in the three situations, respectively.

Recently, Mahesh et al. [8] classified different cardiovascular diseases considering also congestive heart failure and myocardial infarction using multilayer perceptron classifier with several linear and non-linear HRV parameters. The RR segments were taken from the PTB diagnostic ECG database and from the CHF database published in the Physionet website and they found an average accuracy for all types of diseases of 91%.

On the other hand, some authors have applied machine learning techniques to identify IHD patients without considering HRV parameters as features. In particular, Rajeswari et al. [9] created an artificial neural network with an accuracy of 82,2% using only clinical parameters such as body mass index, blood pressure, cholesterol, diabetes and hereditary from 712 patients. In addition, Kukar et al. [10] used diagnostic methods like clinical examinations, exercise ECG testing and myocardial scintigraphy to create a neural network with an accuracy of 92% in a sample of 327 patients.

Since parameters obtained from HRV analysis proved to help reach high performance in the classification of some cardiovascular disease, different from IHD, in this study we applied a specific machine learning technique to parameters coming from HRV analysis. In particular, we evaluate the accuracy of a multi-layer feed forward neural network applied to some HRV parameters extracted from ECG in addition to the left ventricular ejection fraction (LVEF) valued in a non-invasive way.

## II. METHODS

### A. Study population

The study population consisted of 243 subjects consecutively enrolled from January 2012 to December 2012 at the Cardiovascular Department, ASUGI, Trieste. Of these, 156 (80 male and 76 female, aged 53 ± 21) were normal subjects and 87 (72 male and 15 female, aged 71±10) suffered from IHD. IHD was defined by clinical assessment with typical symptoms such as angina and laboratory test (elevation of Troponin I), ECG repolarization abnormalities (T-wave and ST), wall motion abnormalities (hypokinesia or akinesia) with echocardiography, and finally confirmed by coronary angiography. The study was performed according to the Declaration of Helsinki and all patients gave written consent.

### B. HRV parameters

All subjects performed a 24h Holter monitoring session using three channels tracking recorder (Sorin Group, Italy). The ECG signals (Fig. 1 and 2) were sampled at 200 Hz and RR intervals were automatically extracted from records by using SyneScope analysis software. Data were analyzed by using proprietary MATLAB® (MathWorks, USA) program that examined segments of 300s each [11]. The subjects' RR time series were aligned to have a common start time because the original recordings started at different times. Each RR segment was accepted for analysis only if the longest ectopic beats sequence or the longest artefact was shorter than 10s and if the total duration of artefacts and ectopic beats in each segment were shorter than 20% of the segment duration [12]. To obtain a constant sampling time and to replace the identified artifacts and ectopic beats, a cubic spline interpolation based on the normal RR intervals was applied and the new time sequence was resampled at 2 Hz.

Several linear (in time and spectral) and non-linear parameters were evaluated on each segment and subject. The mean value along the 24h was calculated for each parameter and, finally, the average among the subjects of each group was considered. In particular, as linear parameters meanRR (mean of the RR intervals), SDNN (standard deviation of RR intervals), RMSSD (root mean square of the squared differences of successive RR intervals), NN50 (number of differences of successive RR intervals greater than 50ms) and pNN50 (proportion of NN50 divided by the total number of RR intervals) were calculated [13].

The frequency domain was performed by estimating the power spectral density (PSD) of each RR segment by periodogram method, and the power both in the low-frequency (LF, from 0.04 to 0.15Hz) and in the high-frequency (HF, from 0.15 to 0.40Hz) bands as well as the ratio between them (LF/HF ratio) were calculated and averaged along the 24h. These powers were also expressed in normalized units (LFn, HFn), dividing the power of each band by the total power.

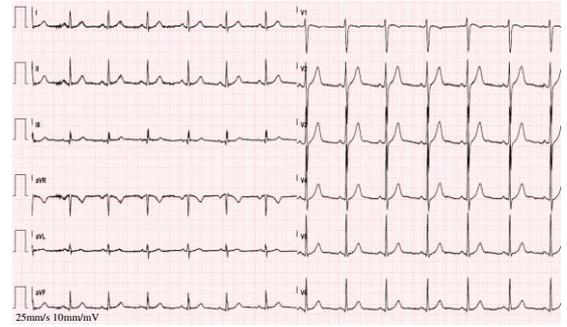

Fig.1. ECG example of normal subject.

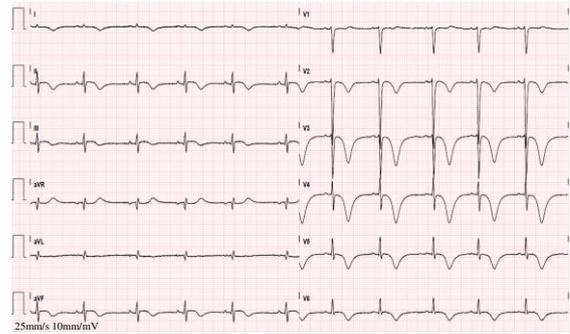

Fig.2. ECG example of IHD patient.

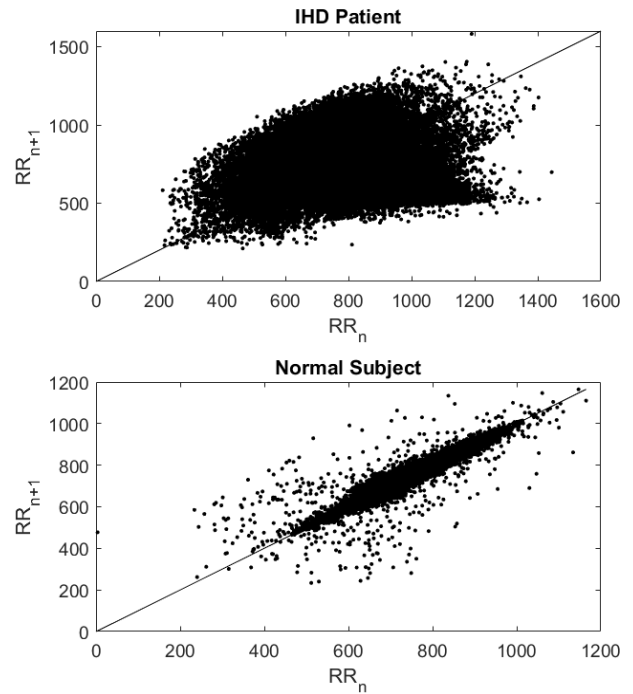

Fig.3. Examples of Poincaré plot in a Normal subject and in an IHD patient.

Non-linear analysis was carried out by evaluating Poincaré parameters, Fractal Dimension (FD) and the beta exponent. The Poincaré plot (Fig. 3) is an alternative representation of a time series and the corresponding parameters (SD1, SD2, SD1/SD2) refer to the short-term and long-term variability of RR sequence [14]. The fractal dimension (FD) quantifies the fractal-like behavior of a time series. It was evaluated by using Higuchi's algorithm [15]. The power-law beta exponent, is an index of both self-affinity and system complexity; it was calculated as the slope of the regression line in the relationship between the ln(PSD) and ln(Frequency) [16].

## C. Neural Network Classifier

In order to classify IHD, we used the HRV parameters together with a clinical parameter, nominally the left ventricular ejection fraction (LVEF), age and gender. A multi-layer feed forward neural network with sigmoid activation function was used. A series of networks with a different number of input nodes (varying between 7 and 15), hidden nodes (between 2 and 10) and two output nodes were tested. The training and test sizes were respectively 75% and 25% of the total number of data.

The training of the neural network ends if the sum of the square errors for all data was less than 0.05 or the maximum number of training epochs (1000 iterations) was reached. The performance of each network classifier was measured (Table I) through accuracy (ACC), sensitivity (SEN), specificity (SPE) and precision (PRE). Finally, the receiver operating characteristics curve (ROC) was used to depict trade-offs between hit rate (sensitivity) and false alarm rate (1.0-specificity) and the area under the curve (AUC) was evaluated.

## III. RESULTS

In order to evaluate which network structure and which combination of parameters presented the highest accuracy, we examined 135 combinations of different network structures varying the number of inputs (equal to the number of the considered parameters) from 4 to 18 and of the hidden nodes between 2 and 10. The LVEF, gender and age parameters were included as the inputs in each network structure considered. A total of 33362 different network combinations was examined and the network with the highest AUC using the validation dataset was selected. The structure of this network presented one hidden layer with 7 nodes and the combination of LVEF, gender, age, SDNN, pNN50, beta exponent and SD2 as inputs. Table 1 shows the performance measurements of the classifier considering training and validation dataset.

TABLE I. CLASSIFICATION PERFORMANCE MEASUREMENTS OF THE CLASSIFIER

| Measures | Formula | Training set performance | Validation set performance |
|---|---|---|---|
| ACC | $\frac{TP+TN}{TP+TN+FP+FN}$ | 98.9% | 82% |
| SEN | $\frac{TP}{TP+FN}$ | 97% | 65% |
| SPE | $\frac{TN}{TN+FP}$ | 100% | 90.2% |
| PRE | $\frac{TP}{TP+FP}$ | 100% | 76.5% |

Fig. 4 reported the confusion matrices with the true positive (TP), the true negative (TN), false positive (FP) and false negative (FN) values and their percentages calculated applying this network to the training and the validation datasets. In the training phase, the accuracy was of 98.9% and the error rate was of 1.1% while using the validation dataset the accuracy was of 82% with an error rate of 18%.

Fig. 5 shows the ROC curves of the ANN performance obtained constructed using the training and the validation dataset. The area under the ROC curve, i.e. the AUC performance value, was 0.99 for the training dataset and 0.83 for the validation dataset.

|  |  | Training Confusion matrix | | Validation Confusion matrix | |
|---|---|---|---|---|---|
|  |  | Target Classes | | Target Classes | |
|  |  | IHD | NORMAL | IHD | NORMAL |
| Output classes | IHD | TP 65 97% | FP 0 0% | TP 13 65% | FP 4 9.8% |
| Output classes | NORMAL | FN 2 3% | TN 115 100% | FN 7 35% | TN 37 90.2% |

Fig.4. Confusion matrices of the Training and Validation datasets.

## IV. DISCUSSION

The use of ANN applied to linear (time and frequency domains) and non-linear measures of HRV for the classification of some cardiac diseases, such as congestive heart failure and myocardial infarction, has been demonstrated to have a high accuracy up to 91% [6], [7], [8].

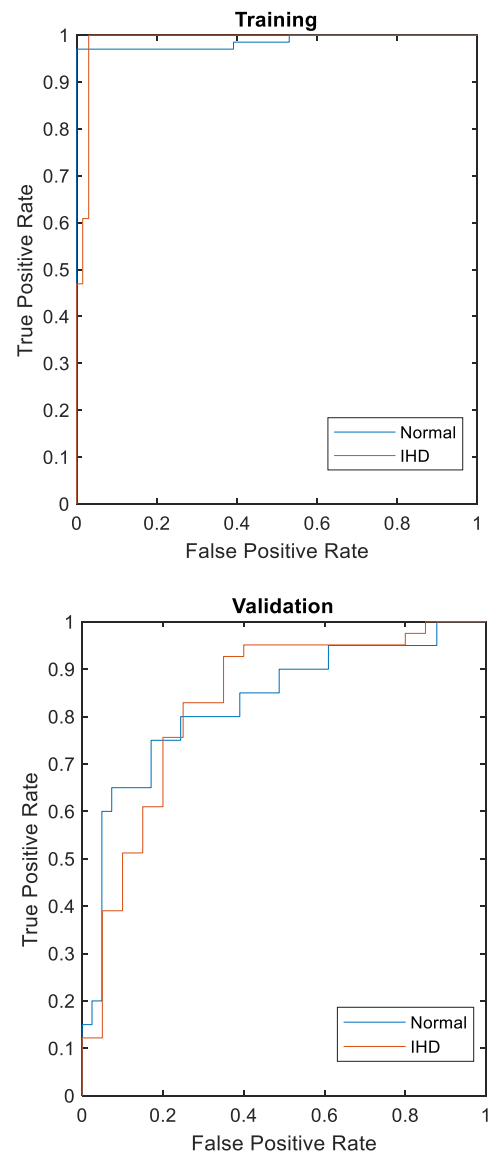

Fig.5. Receiver Operating Characteristic (ROC) Curves of the Artificial Neural Networks in training (top panel) and validation (bottom panel) datasets. class 1: Normal subjects class 2: IHD patients.

However, the classification based on HRV parameters has not yet been used to identify IHD patients. Indeed, IHD's diagnosis is based on clinical features and some non-invasive diagnostic test such as ECG, Exercise ECG, functional imaging test (stress echocardiography, coronary CTA, myocardial scintigraphy), but invasive angiography is used for confirmation of the diagnosis in patients with an uncertain diagnosis on non-invasive testing [9], [10].

On the other hand, patients suffered from IHD showed significant different values of HRV parameters than normal subjects that could be explained by an increased sympathetic tone and/or a decreased parasympathetic tone [4], [5].

In this paper, we had evaluated the performance of an artificial neural network trained and tested by using only HRV parameters together with a clinical parameter (LFEV) measured in a non-invasive way. Among the several combinations of the ANN examined, the best network allowed to obtain a high training performance (99%) and a good validation accuracy (83%).

With the validation test, our study revealed that the ANN modelling was more specific than sensitive tending to misclassify more IHD patients as normal, confirming the results obtained during the training. During the training, no normal subject was classified as IHD patient while, during the validation test, about 10% of the normal subjects were identified as suffering from IHD. However, the difference between the percentage of false positive and of false negative, negligible during the training, was instead large during the validation test, implying that the ANN modelling was not able to balance sensitivity and specificity. This could be due to the low number of IHD subjects during the validation phase.

Improvement in modelling performance may be provided by increasing the number of subjects, especially IHD patients, with the aim of making the model more representative. Moreover, a further improvement in accuracy could be expected by using other machine learning techniques such as decision tree or single vector machine.

## Acknowledgment

Work partially supported by Master in Clinical Engineering, University of Trieste.